\documentclass[preprint, showpacs,preprintnumbers,amsmath,amssymb,nofootinbib]{revtex4}
\usepackage{mathrsfs}
\usepackage{amsmath}
\usepackage{amsfonts}
\usepackage{latexsym}
\usepackage{amsfonts}
\usepackage{graphicx}
\usepackage{epsf}
\usepackage{dcolumn}
\usepackage{bm}

\newcommand{\bea}[1]{\begin{eqnarray}\label{#1}}
\newcommand{\eea}{\end{eqnarray}}

\def\gsim{ \lower .75ex \hbox{$\sim$} \llap{\raise .27ex \hbox{$>$}} }
\def\lsim{ \lower .75ex \hbox{$\sim$} \llap{\raise .27ex \hbox{$<$}} }

\begin{document}
 \title{Is the  present cosmic expansion decelerating?  }

\author{Puxun Wu and  Hongwei Yu }
\address
{  Center for Nonlinear Science and Department of Physics, Ningbo
University,  Ningbo, Zhejiang 315211, China
%$^2$ Department of Physics and Key Laboratory of Low Dimensional
%Quantum Structures and Quantum Control of Ministry of Education,
%Hunan Normal University, Changsha, Hunan 410081, China
}

\begin{abstract}

We probe the recent cosmic expansion by directly reconstructing the
deceleration parameter $q(z)$ at recent times with a linear
expansion at $z=0$ using the low redshift SNIa and BAO data. Our
results show that  the observations seem to favor a slowing down of
the present cosmic acceleration.  Using only very low redshift SNIa
data, for example, those within $z<0.1$ or $0.2$, we find that our
Universe may have already entered a decelerating expansion era since
a positive $q(0)$ seems to be favored. This result is further
supported by  a different approach which aims to reconstruct $q(z)$
in the whole redshift region. So, the accelerating cosmic expansion
may be just a transient phenomenon.

\end{abstract}
\maketitle

\section{Introduction}\label{sec1}

The fact that our Universe entered a phase of  accelerating
expansion at redshift $z$ less than $\sim 0.5$ is well established
by several data sets~\cite{SNIa, BAO, CMB}, and most analysis seem
to suggest that this cosmic  acceleration is increasing with time.
However, recently,  Shafieloo {\it et al.}~\cite{slowing} found, by
using the Constitution type Ia supernova data
(SNIa)~\cite{constitution} and data from the baryon acoustic
oscillation (BAO) distance ratio of the distance measurements
obtained at $z = 0.2$ and $z = 0.35$ in the galaxy power
spectrum~\cite{bao1, bao2}, and the CPL parametrization~\cite{CPL}
for the equation of state for dark energy, that the acceleration  of
the cosmic expansion is probably slowing down. At the same time,
they also found that this result is dependent both on the data and
the parametrization used. For example, they showed that observations
still favor an increasing  cosmic acceleration when BAO and SNIa is
combined with  the cosmic microwave background radiation (CMB) data
from WMAP7~\cite{Komatsu2010}. However,  if a different
parametrization or a subsample (SNLS+ESSENCE+CfA) of the
Constitution SNIa is used, both SNIa+BAO and SNIa+BAO+CMB favor that
the cosmic acceleration is slowing down. Thus, two different, even
opposite, results have been obtained. The discrepancy may arise
because of either the systematics in some data or that the CPL
parametrization is not versatile-enough to accommodate the evolution
of dark energy implied by the data. The same issue was also studied,
recently, by Gong {\it et al.}~\cite{gong} and Li {\it et
al.}~\cite{Li2010, Li20102}. They found that the systematics in data
sets, the parametrization of dark energy as well as the system error
in SNIa all affect outcome for the reconstructed cosmic expansion
history. So, up to now, we can not answer for sure the question as
to whether the current cosmic acceleration is slowing down or
speeding up.  One of the main difficulties is that the evolutionary
properties of dark energy is still unknown.

In the present paper, we take a different approach by directly
reconstructing the evolutional behavior of the deceleration
parameter $q(z)$ at recent times using observational data without
any assumption on the cause of the dynamical evolution of the
Universe, whether it be dark energy or modified gravity. Since we
are only interested in the property of the current cosmic evolution,
we use a linear expansion for  $q(z)$, i.e., we let $q(z)=q_0+q_1 z$
\footnote{This  was  firstly proposed  in Ref.~\cite{Turner2002} to
probe the cosmic evolution from the SNIa data.}, which should be a
very reasonable approximation in the low redshift regions, for
example, $z<0.2$.  Because the linear expansion may only be valid in
the low redshifts, we only use the low redshift data, such as the
SNIa data points in the low redshift regions and BAO, to determine
the parameters $q_0$ and $q_1$ to obtain the evolutionary behavior
of $q(z)$ at $z\sim 0$, which may give a qualitative result for the
present cosmic acceleration.

\section{data and results}
The data sets used here include the SNIa and BAO. For SNIa, the
latest Union2 compilation released by the Supernova Cosmology
Project (SCP) collaboration recently~\cite{Union2} is considered
since it is the currently largest published SNIa sample. This Union2
consists of 557 data points in the range $0.015<z<1.4$. But we only
select data points at low redshifts. To give a comparison of
different cases, we consider four kinds of low redshift regions,
i.e.,  $z\leq 0.1$, $z\leq 0.2$, $z\leq 0.35$ and $z\leq 0.5$. For
$z\leq 0.1$, $0.2$, $0.35$ and $0.50$, there are $166$, $220$,
$318$, and $402$ data points, respectively.

For BAO data, as in Refs.~\cite{slowing,gong,Li2010}, we use the
distance ratio obtained at $z=0.20$ and $z=0.35$ from the joint
analysis of the 2dF Galaxy Redsihft Survey and SDSS
data~\cite{bao2}. So, only in the cases of $z\leq 0.35$ and $z\leq
0.5$, we can combine the SNIa and BAO to probe the evolutionary
behavior of $q(z)$.

The results are shown in Fig.~(1).  The upper left, upper right,
down left and down right panels show the results from the SNIa with
$z\leq 0.10$, $0.20$, $0.35$ and $0.50$, respectively. The regions
between red dashed lines represent the allowed evolutionary behavior
from the Union2 SNIa data at the $1\sigma$ confidence level, while
the green regions are the results from SNIa+BAO. Apparently, for all
cases, the Union2 SNIa seems to favor that the present cosmic
acceleration is slowing down since the best fit line of the
deceleration parameter is increasing with the decreasing of $z$,
although at $1\sigma$ confidence level the case of an increasing
cosmic acceleration cannot be ruled out. With the addition of BAO
data, the decreasing trend of the cosmic acceleration becomes more
evident. For the cases $z\leq 0.10$ and $0.2$, the observations
 not only favor a slowing down of the cosmic acceleration, but also seems to indicate that
the Universe  has probably entered  a decelerating expansion at the
present since $q(0)$ is largely in the positive region. This
behavior is more evident when data points within $z<0.1$ are used to
reconstruct $q(z)$ than that within $z<0.2$. So, our result seems to
suggest  that the accelerating expansion of our Universe might be
just a transient phenomenon. It is interesting to note that a
similar result was also obtained in Ref.~\cite{Guimaraes2010} with a
cosmographic method.

It has been pointed out that the SNIa data obtained from different
light curve fitting methods, such as SALT-II and MLCS2k2, may give
different results on the property of dark energy~\cite{Li20102,
Bengochea2010, Kessler2009}. Since the present Union2 SNIa set is
only analyzed with the SALT2 light curve fitter, we now carry out a
discussion of other SNIa sets, such as
Constitution~\cite{Hicken2009} and SDSS-II~\cite{Kessler2009}, and
consider the effect of the different light curve fitters. The best
fit results are shown in Tab.(1) for a linear expansion,
$q(z)=q_0+q_1 z$,  using the SNIa data points within $z\leq 0.2$. It
is easy to see that, except for the case of SDSS-II with the MCLS2k2
fitter, all other data sets seem to favor a slowing down of the
cosmic acceleration because $q_1<0$ and a present decelerating
cosmic expansion since $q_0>0$.

\begin{table}[!h] \tabcolsep 3pt \caption{\label{Tab1} Summary of  the
best fit values for $q(z)=q_0+q_1 z$ from the Constitution and
SDSS-II SNIa data sets with $z\leq 0.2$. } \vspace*{-2pt}
\begin{center}
\begin{tabular}{|c|c|c|c|c|c|}
  \hline
              & $q_0$   & $q_1$            \\ \hline
 $SDSS-II (MCLS2k2)$  & $-1.23$ & $11.76$  \\ \hline
 $SDSS-II (SALT2)$  & $0.389$ & $-15.25$    \\ \hline
 $Constitution(MLCS2k2)$  & $1.556$ & $-48.21$   \\ \hline
 $Constitution(SALT2)$  & $1.107$ & $-39.94$   \\ \hline
\end{tabular}
       \end{center}
       \end{table}

Finally, to get more complete picture for the evolution of the
cosmic acceleration, let us try another  different approach, which
aims to reconstruct $q(z)$ with observational data in all redshifts.
Now, we divide the whole redshift region into five segments, as
shown below, and  assume that  the value of the deceleration
parameter is a constant in each segment:
\begin{eqnarray}
 z:&&\quad 0-0.05\quad 0.05-0.2\quad 0.2-0.5\quad 0.5-1.0\quad 1.0-\\
\nonumber
 q:&&\quad q_0\quad \quad\quad \quad   q_1\; \quad \quad\quad \quad
q_2\quad \quad\quad \quad   q_3\quad \quad\quad \quad   q_4
\end{eqnarray}
We then use the observational data to constrain these five free
parameters. Besides the Union2 SNIa and BAO data, the CMB shift
parameter from WMAP7~\cite{Komatsu2010} is also added in our
analysis. The results are shown in Fig.~(\ref{Fig3}). In this
figure, the blue solid, green solid and red dashed lines represent
the best fit results obtained from SNIa, SNIa+BAO and SNIa+BAO+CMB,
respectively. The yellow region is the $1\sigma$ confidence level
from SNIa+BAO+CMB. The best fit results show that all the
observational data favor that  the cosmic expansion is decelerating
in both the high and very low redshift regions, which means that the
accelerating cosmic expansion is possibly a transient phenomenon,
although at the $1\sigma$ confidence level the observations still
allow the possibility of a currently accelerating cosmic expansion.
In addition, we find that the SNIa+BAO and SNIa+BAO+CMB give the
consistent results. So, the tension between low redshift data (SNIa
and BAO) and high redshift one (CMB) found in Refs.~\cite{slowing,
Li2010, Jassal2005} disappears.

\section{conclusion}
In summary, we have probed the recent cosmic expansion by
reconstructing the deceleration parameter $q(z)$ with a linear
expansion  at $z=0$ using the low redshift Union2 SNIa data and BAO
data. We find that  the observations seem to favor a slowing down of
the present cosmic acceleration.  Using only very low redshift SNIa
data, for example, those in $z<0.1$ and $0.2$, we obtain that our
Universe may have already entered  a decelerating expansion era at
the present since a positive $q(0)$ seems to be favored. This means
that the accelerating cosmic expansion is probably a  transient
phenomenon. To see  the effect of light curve fitting  method on our
results, we also consider the Constitution and SDSS-II SNIa datasets
with the SALT2 and MLCS2k2 light curve fitters, respectively. The
best fit result shows that except for the SDSS-II (MLCS2k2), all
other data favor that the cosmic acceleration is possibly  slowing
down and the present cosmic expansion may be decelerating.
Furthermore, we also tackled the issue by dividing the whole
redshifts into five segments, assuming $q(z)$ be a constant in each
segment and then fitting the data from SNIa+BAO+CMB, and  found
that a transient
 accelerating cosmic expansion
 is plausible. Finally, we must point out that  a
currently accelerating cosmic expansion cannot be ruled out at the
$1\sigma$ confidence level although the best fit results do not
favor it. To obtain a clearer answer, we still need to wait for more
data.

\begin{figure}[htbp]
 \centering
\includegraphics[width=6cm]{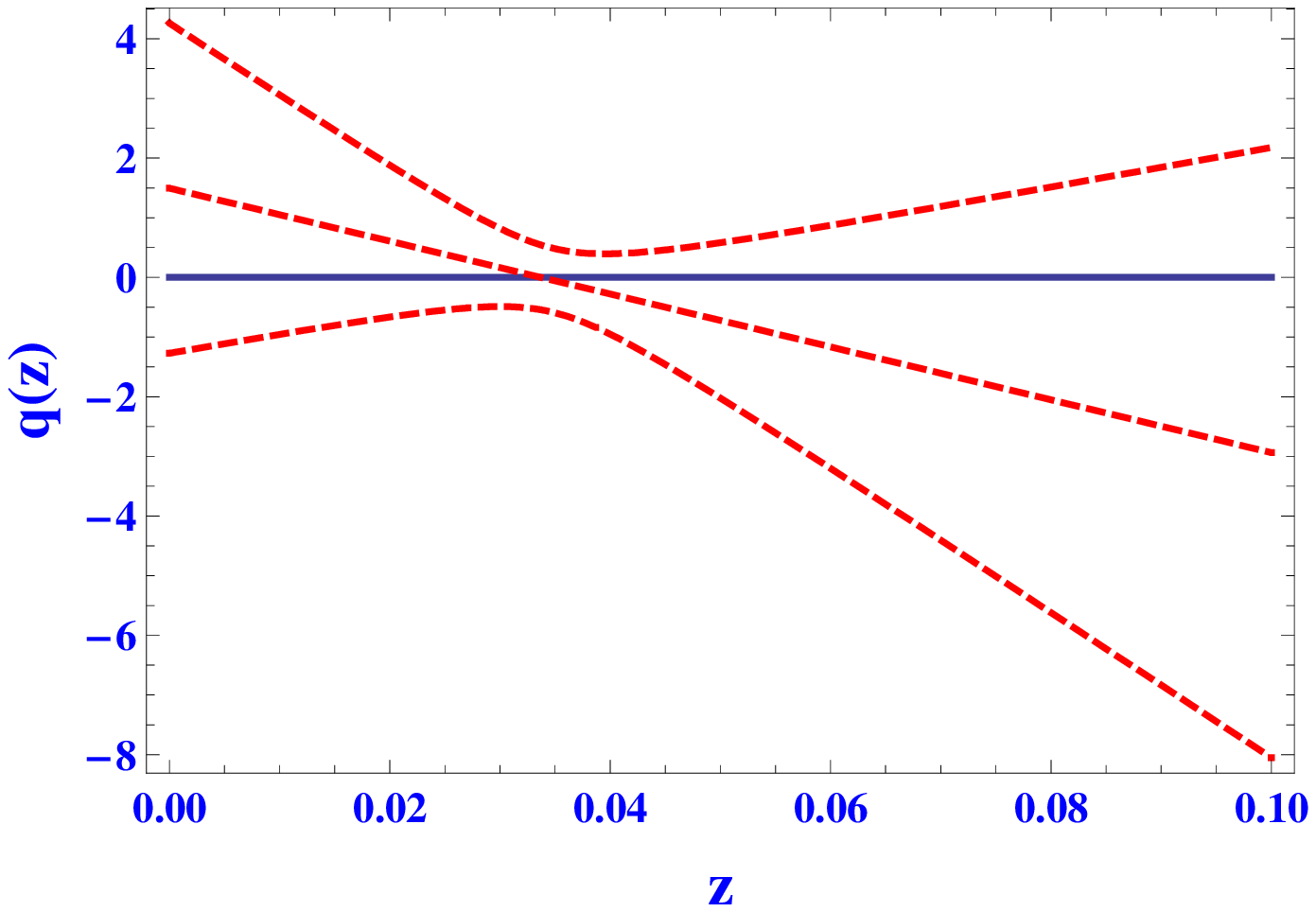}\quad\includegraphics[width=6cm]{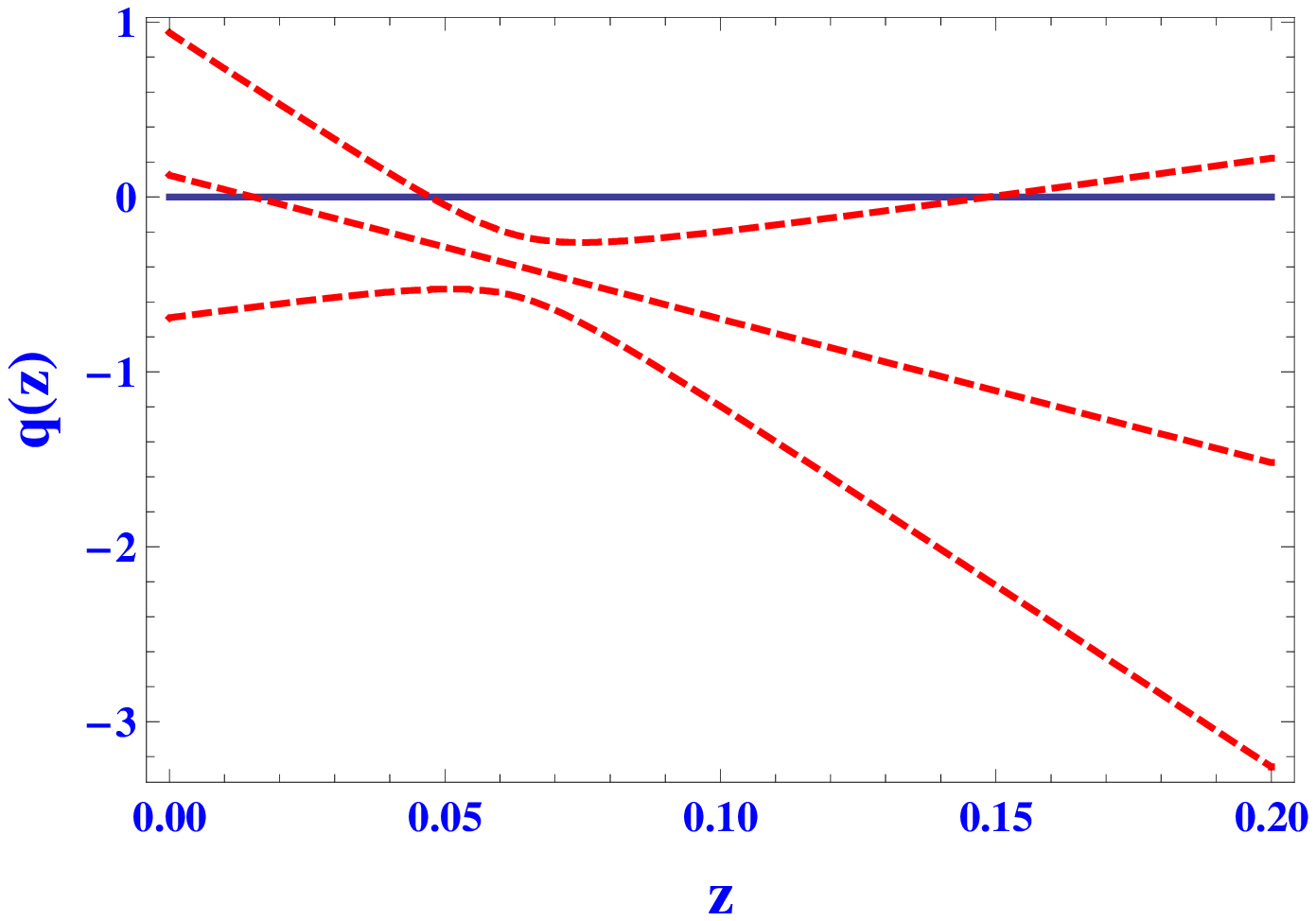}\quad\includegraphics[width=6cm]{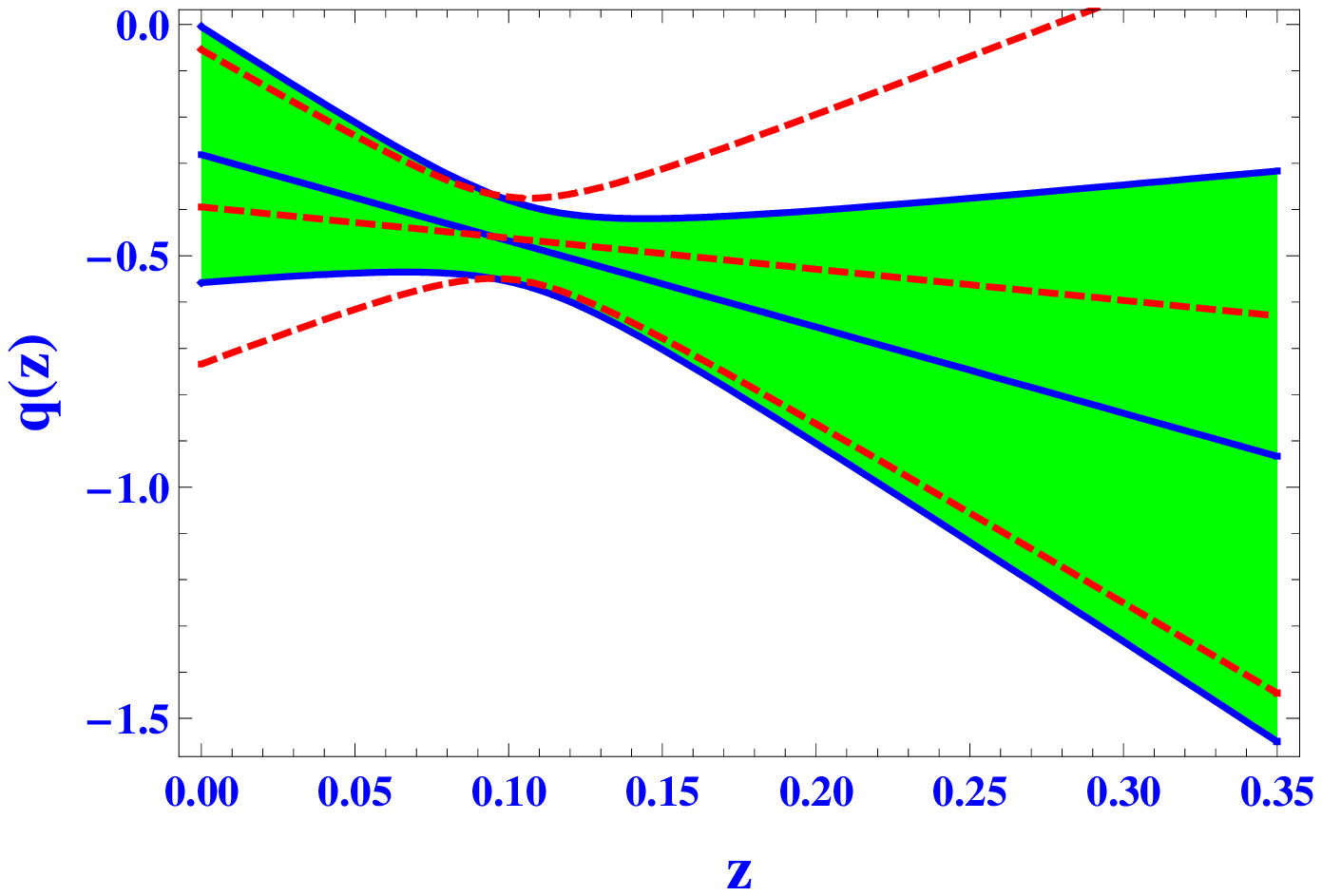}\quad
\includegraphics[width=6cm]{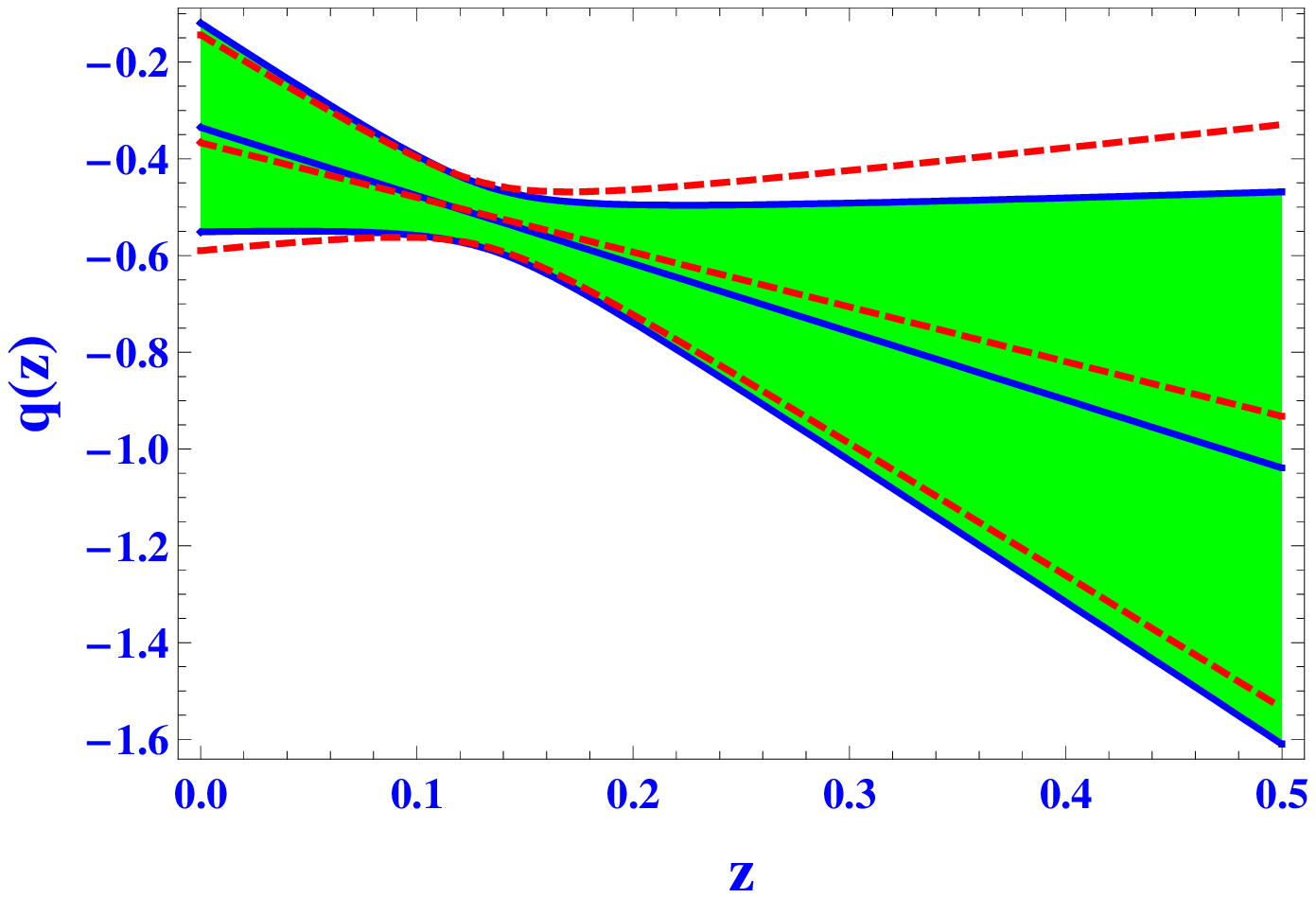}\quad
\caption{\label{Fig1}  The evolutionary behavior of $q(z)$ with
$q(z)=q_0+q_1 z$. The red dashed lines are the results form the
Union2 SNIa and the green regions are that from SNIa+BAO at the
$1\sigma$ confidence level. The upper left, upper right, down left
and down right panels correspond to the SNIa with $z\leq 0.1$,
$0.2$, $0.35$ and $0.5$, respectively.    }
 \end{figure}

\begin{figure}[htbp]
 \centering
 \includegraphics[width=9cm]{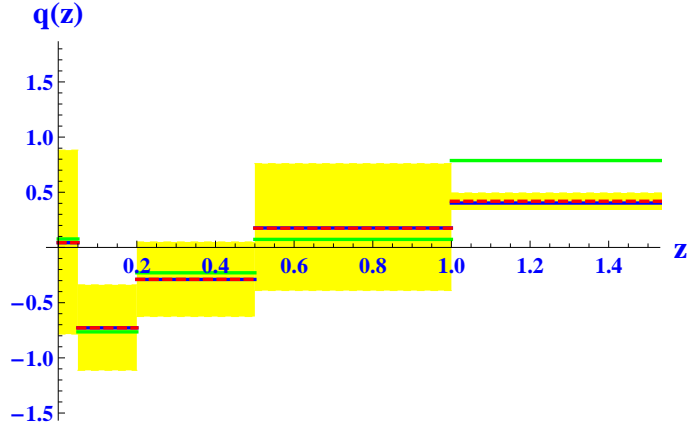}
\caption{\label{Fig3}  The evolutionary behavior of $q(z)$ for model
given in Eq.~(1). The blue solid, green solid and red dashed lines
represent the best fit results form Union2 SNIa, SNIa+BAO and
SNIa+BAO+CMB, respectively. The yellow regions are the $1\sigma$
confidence level from SNIa+BAO+CMB.  }
 \end{figure}

\begin{acknowledgments}

We would like to think Prof. Chao-Qiang Geng and members of his
group for their very helpful discussions and NTHU in Taiwan for
hospitality where part of this work was done. This work was
supported in part by the National Natural Science Foundation of
China under Grants Nos. 10775050, 10705055, 10935013 and 11075083,
Zhejiang Provincial Natural Science Foundation of China under Grant
No. Z6100077, the FANEDD under Grant No. 200922, the National Basic
Research Program of China under Grant No. 2010CB832803, the NCET
under Grant No. 09-0144, the PCSIRT under Grant No. IRT0964, and
K.C. Wong Magna Fund in Ningbo University.

\end{acknowledgments}

\end{document}